\documentclass[reprint,showpacs,preprintnumbers,amsmath,amssymb,prb,longbibliography]{revtex4-1}
\usepackage{graphicx}
\usepackage{dcolumn}
\usepackage{bm}
\usepackage{epsfig}
\usepackage{color}
\usepackage{ulem}

\begin{document}


\title{How branching can change the conductance of ballistic semiconductor devices}

\author{D.~Maryenko}
\email{maryenko@riken.jp}
\altaffiliation[Current address: ]{Correlated Electron Research Group (CERG), RIKEN Advanced Science Institute, Wako 351-0198, Japan}
\affiliation{Max Planck Institute for Solid State Research, D-70569 Stuttgart, Germany}
\author{J.~J.~Metzger}
\affiliation{Max Planck Institute for Dynamics and Self-Organization, D-37077 G\"{o}ttingen, Germany}
\author{F.~Ospald}
\affiliation{Max Planck Institute for Solid State Research, D-70569 Stuttgart, Germany}
\author{R.~Fleischmann}
\affiliation{Max Planck Institute for Dynamics and Self-Organization, D-37077 G\"{o}ttingen, Germany}
\author{V.~Umansky}
\affiliation{Braun Center for Submicron Research, Weizmann Institute, Rehovot, 76100, Israel }
\author{T.~Geisel}
\affiliation{Max Planck Institute for Dynamics and Self-Organization, D-37077 G\"{o}ttingen, Germany}

\author{K.~v.~Klitzing}
\author{J.~H.~Smet}
\affiliation{Max Planck Institute for Solid State Research, D-70569 Stuttgart, Germany}


\begin{abstract}
We demonstrate that branching of the electron flow in semiconductor nanostructures can strongly affect macroscopic transport quantities and can significantly change their dependence on external parameters compared to the ideal ballistic case even when the system size is much smaller than the mean free path. In a corner-shaped ballistic device based on a GaAs/AlGaAs two-dimensional electron gas we observe a splitting of the commensurability peaks in the magnetoresistance curve. We show that a model which includes a random disorder potential of the two-dimensional electron gas can account for the random splitting of the peaks that result from the collimation of the electron beam. The shape of the splitting depends on the particular realization of the disorder potential. At the same time magnetic focusing peaks are largely unaffected by the disorder potential.
\end{abstract}


\maketitle

\section{Introduction}

Transport of two-dimensional electrons in state-of-the-art modulation-doped semiconductor heterostructures suffers mainly from small angle  scattering off charge fluctuations in the donor layer, which is separated from the two-dimensional electron system by a spacer layer \cite{LL5, LL6, LL7}. The amplitude of the disorder is only a few percent of the Fermi energy, and the electron mean free path can exceed hundreds of microns \cite{LL8, LL9}. Under these circumstances, transport through nanostructures much smaller than the mean free path is assumed ballistic, i.e. individual electrons follow almost perfectly the paths prescribed by Newton's law under the external applied forces and the forces associated with the nanostructure's confinement potential. Impurity scattering is thought of as negligible. Nevertheless, experiments, which used spatially resolved recordings of the change in conductance induced by a charged tip of a scanning probe microscope, showed that the current density emerging from a quantum point contact was branched as a result of small angle scattering only \cite{LL1,LL2,LL3}. The reported conductance changes in these and further experiments using the same measurement techniques were much smaller than the conductance quantum $e^2/h$. At the same time many experiments have been interpreted in terms of purely ballistic effects, e.g. focusing by electrostatic \cite{LL10} and magnetic \cite{L6} fields or wall geometries \cite{LL12} even in semiconductor materials with shorter mean free paths. It could thus seem that branching has little impact on macroscopic transport quantities and hence is of no relevance for most transport experiments.

\begin{figure}
\includegraphics{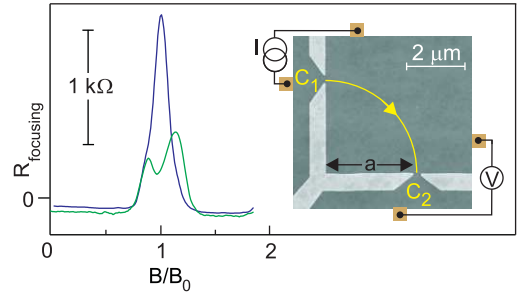}
\caption{\label{fig:0} The first magnetoresistance peaks recorded for two mesoscopic devices fabricated from the same heterostructure. The inset shows the scanning electron microscope picture of the novel magnetic focusing device. $B_0$ is the magnetic field at which the cyclotron radius is commensurate with the wall length of the device.}
\end{figure}

\begin{figure}
\includegraphics{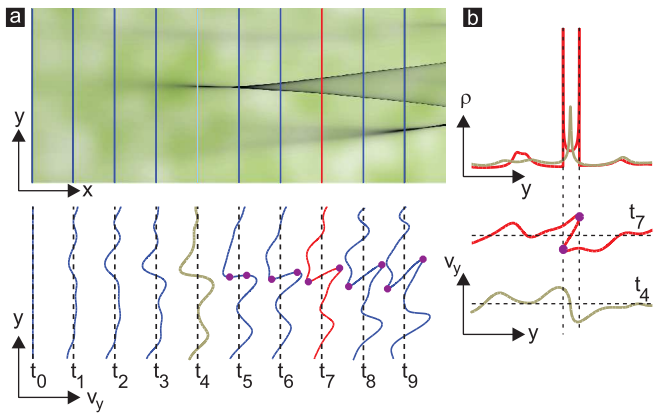}
\caption{\label{fig:1}  Branching of a plane wave front propagating in x-direction. (a) The electron wave flow intensity approximated by classical ray dynamics is plotted in gray scale as the plane wave propagates along the $x$-direction in the disorder landscape shown as a green/white color rendition in the background. The force in $x$-direction is ignored (quasi-2D). The wave front remains a vertical line in coordinate space. Vertical lines mark the position of the plane wave front at times $t_1$, $t_2$, $\ldots$,$t_9$. The bottom panel displays the wave fronts at these times in phase space ($y$,$v_y$). Caustics are identified at the turning points (purple dots). (b) Flow density $\rho$ as a function of $y$ at times $t_4$ and $t_7$ (vertical cross-sections of the flow density plot in a). Also shown at the bottom are the corresponding wave fronts in phase space. The flow density peaks at the caustics.}
\end{figure}
\begin{figure}[!t]
\includegraphics{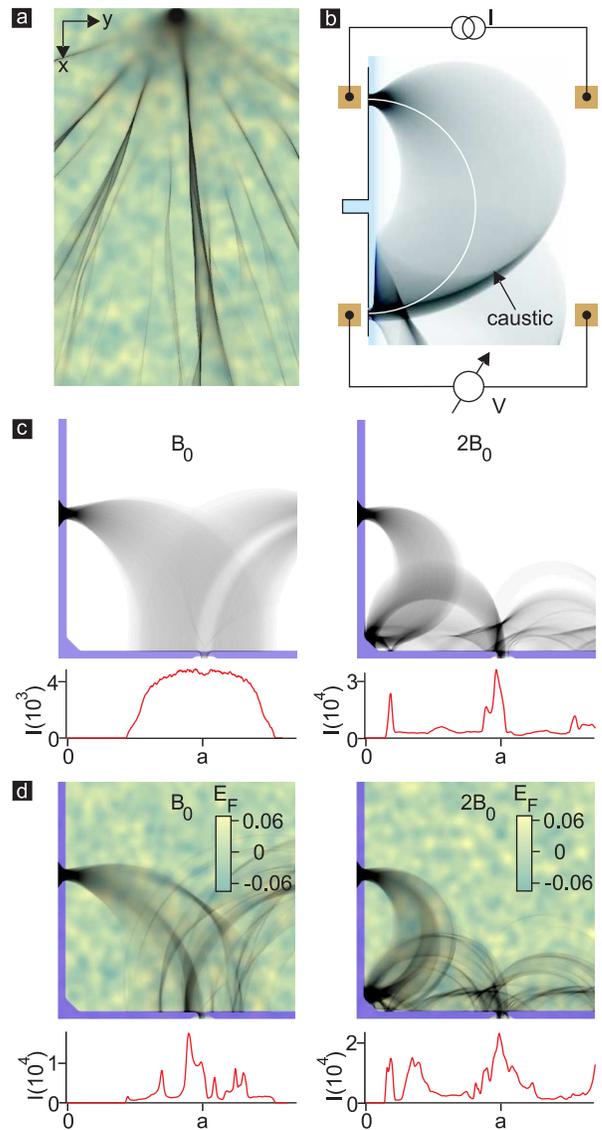}
\caption{\label{fig:2} Influence of disorder on the flow density $\rho$ (gray scale rendition) emitted from a point source. (a) gray scale rendition of the flow density $\rho$ in the $(x,y)$-plane when particles are emitted from a point source with a cosinusoidal angular distribution. They are subjected to the disorder potential color-coded in green and white. Forces in both x and y-direction were taken into account and the magnetic field is absent. (b) The formation of caustics in a transverse magnetic focusing geometry. Particles are emitted from the top point contact. The flow density is plotted on a gray scale at a magnetic field where a caustic forms at the collecting point contact. (c) Collimation (left panel) at $B=B_0$ and magnetic focusing at $B=2B_0$ (right panel) in a corner device consisting of two quantum point contacts placed at a 90$^\circ$ angle.  Red curves in the lower panels show the flow density hitting the lower wall. (d) Same calculations as in c but in the presence of weak disorder. The standard deviation of the amplitude of the disorder potential corresponds to 2$\% E_F$.} 
\end{figure}


In this article, however, we use a novel magnetic focusing device to demonstrate both experimentally and theoretically that the macroscopic transport quantities of nanostructures can in fact be strongly influenced by branching. This poses the question of how ballistic {\it transport effects} observed in many supposedly ballistic experiments actually are. To illustrate this let us examine the two sections of magnetoresistance curves from two different samples fabricated to the exact same specifications shown in Fig.~\ref{fig:0} (details will be given below). In an ideal (i.e. ballistic) sample patterned in this specific layout (depicted in the inset) electrons emitted from point contact C$_1$ are deflected by a magnetic field and for a certain value $B=B_0$ will be directed towards the second point contact C$_2$. Therefore, one expects to observe a single peak in an appropriate transport quantity at $B/B_0=1$. Sample 1 shows exactly this behavior, sample 2, however, shows an unexpected splitting of the peak. So far, many experimentalists would argue that sample 2 is a defective sample, where an unfavorably located impurity spoils the measurement and thus that the sample should be discarded. We show, however, that curve 1 and 2 are both fully compatible with exactly the same amount of impurity scattering and that it can not be argued that sample 1 is in any way better than sample 2. On the contrary, even though both samples are actually extremely clean devices, it is a rather fortunate coincidence that curve 1 agrees well with the expectations for an ideal system. In addition, our new magnetic focusing device allows us to explain why these consequences of branching have not been seen in previous magnetic focusing geometries.

\section{Caustic formation by disorder and a magnetic field}
\subsection{Branching of an initially plane wave}
We prelude the detailed presentation of our results by an illustrative explanation of the mechanism of branch formation, as depicted in Fig.~\ref{fig:1}a and b. An initially homogeneous particle flow or plane wave front (restricted source in momentum space) is propagated along the $x$-direction through a two-dimensional disorder potential. A color rendition (from green to white) of the disorder potential is plotted in the background of panel a. For the sake of simplicity, only the electrostatic force exerted by the disorder potential in the $y$-direction is considered. The force in $x$-direction is ignored, so that the longitudinal velocity $v_x$ stays constant and the wave front remains a vertical line in coordinate space. This model makes the principles of branching particularly easy to understand and yet captures all of its important features \cite{L13}. Later on, an extension of the model will also allow us to study the statistics of the formation of branches in a magnetic field analytically. The particle or flow density  $\rho$ is shown in panel \ref{fig:1}a using a gray scale. While $\rho$ is initially homogeneous or independent of $y$, it develops features at later times as illustrated in panel b where $\rho(y)$ is plotted at selected times. At a time between $t_4$ and $t_5$, a strong peak develops in $\rho$. This heralds the first branch. In the phase space ($y$,$v_y$), the wave front develops a pair of initially coalescing turning points, which subsequently separate  (panel a bottom). Here, the classical ray density in coordinate space diverges and also the quantum mechanical wave intensity would peak nearby. In between the turning points, the wave front folds in phase space and covers the same spatial coordinate three times. As a result, the local density is enhanced. The path traced by a turning point constitutes a fold line or caustic. A branch is referred to as the spatial region in between two such random caustics.

\subsection{Branching and magnetic focusing with a point source}
Figure~\ref{fig:2}a illustrates another instance of caustic formation and branching when particles are emitted from a point source (restricted source in coordinate space). In this example, no approximation concerning the electrostatic force associated with the disorder potential is made and the full two-dimensional particle dynamics is considered.  Branches appear on similar length scales as for the simplified plane wave case due to the same basic mechanism \footnote{See Appendix A}. This setup can be implemented straightforwardly in a GaAs based 2DEG on which a quantum point contact (QPC) is patterned either by etching or the split gate technique \cite{L14,L15,L16,L17,L18,L19}.
\begin{figure*}[!t]
\includegraphics{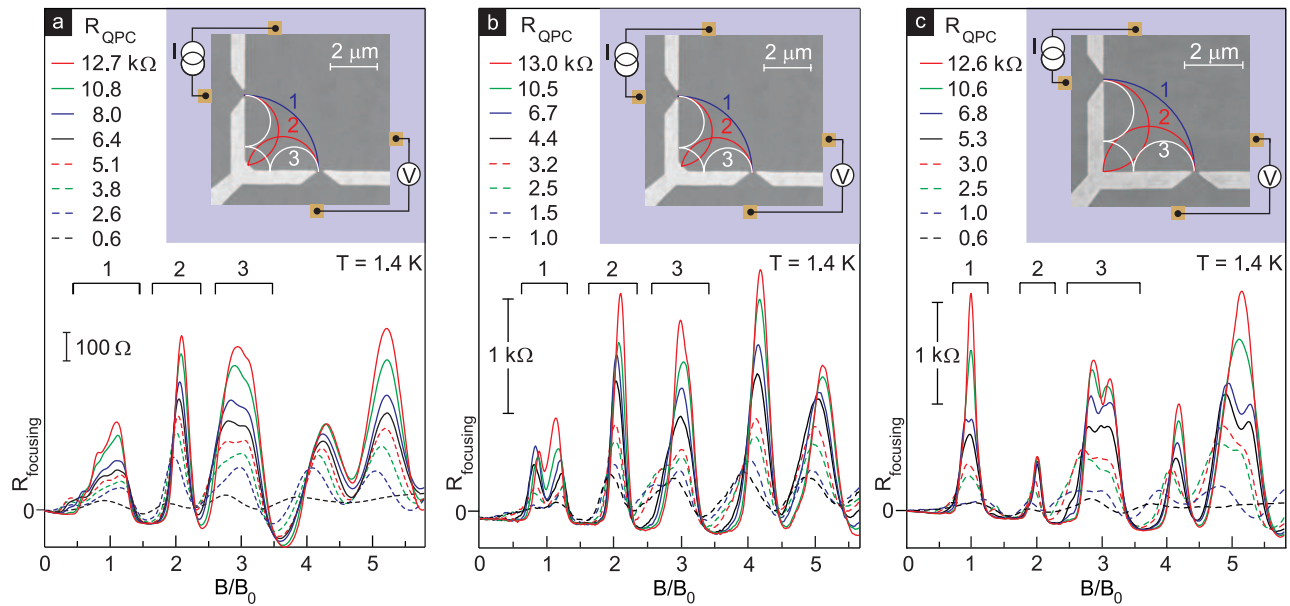}\caption{\label{fig:3}
Magnetoresistance traces measured in corner devices with different QPC resistances, $R_\text{QPC}$. The electron mean free path of the 2DEG is 45~$\mu$m. (a-b) Magnetoresistance traces for two devices with a chamfered corner fabricated from the same wafer show different behavior of the resistance $R_\text{focusing}$ feature near $B/B_0$=1. In each panel, curves are shown for various values of the resistance of the quantum point contact, $R_\text{QPC}$.(c) Magnetoresistance data for a device with a sharp corner design.}
\end{figure*}

It is instructive to oppose the appearance of branches due to the disorder induced formation of random caustics to the focusing of two-dimensional electrons emitted from a point source in a perpendicular magnetic field $B$ in the absence of disorder. We do this, because also in the magnetic focusing problem caustics play an important role \cite{L6,LL3,L21}. In a magnetic field, the electrons execute circular cyclotron orbits with a radius $r=\frac{\hbar\sqrt{2\pi n}}{eB}$ , where $n$ is the electron density. When emitted from a point source, the electron trajectories converge at a distance of one cyclotron diameter away from the point source and a caustic forms. The enhanced local current density can be detected with the help of a second collecting quantum point contact in the transverse magnetic focusing geometry shown in Fig.~\ref{fig:2}b \cite{L6,L22,L23}. The collecting QPC is at a distance $a$ from the emitting contact. The enhanced local density due to magnetic focusing arises at a magnetic field for which the cyclotron diameter $2r$ equals $a$. When enforcing zero net current flow through this QPC and measuring the voltage drop across, the enhanced local density can be detected as a voltage or resistance peak \cite{L6}. One may anticipate that disorder induced branching also produces resistance peaks when sweeping the magnetic field.

\subsection{Focusing in the corner device}
In order to search for evidence of branching in dc ballistic transport we have chosen the corner shaped device depicted in Fig.~\ref{fig:2}c, which, as we will show, can distinguish between deterministic magnetic focusing at the collecting QPC and random focusing caused by branching. The left panel shows trajectories in the absence of disorder at the field $B_0$  for which $r=a$. The flow density along the bottom boundary is plotted and forms a broad peak centered around the collecting point contact opening for this magnetic field. It is not caused by magnetic focusing since a caustic has not yet developed. Rather it results from the collimating properties of the emitting point contact from which electron trajectories leave with a cosinusoidal angular distribution. One may anticipate that it produces a first resistance peak \cite{LL12,L24}. At the field $2B_0$ (right panel of Figure~\ref{fig:2}c), the particle flow converges first into the device corner where a caustic has formed due to magnetic focusing. The corner was chamfered to have well defined reflection. After specular reflection the electron trajectories refocus at the collecting QPC giving rise to another resistance maximum.

Figures~\ref{fig:2}c and d illustrate how the disorder potential affects the particle flow for both values of the magnetic field. While at $2B_0$ the influence of disorder is weak, the flow density has been drastically altered for $B=B_0$ and now exhibits multiple maxima of comparable size. The magnetic focusing feature at $2B_0$ remains largely unaltered and is more robust against disorder induced branching. These conclusions can be generalized and also hold at higher magnetic fields. In the absence of disorder, the flow density reaches a maximum at the collecting QPC for $B= kB_0$, where $k$=1,2,$\ldots$. Collimation and focusing are responsible for these maxima at odd and even values of $k$ respectively. The collimation features are prone to disorder induced splitting, while the focusing features are generally more resilient. As opposed to the conventional transverse magnetic focusing geometry \cite{L6}, in this corner device deterministic focusing and collimation are separated on the magnetic field axis and hence this geometry lends itself particularly well to confirm the theoretical predictions.
\section{Experimental results}
The devices are fabricated from a modulation doped GaAs/AlGaAs heterostructure in which the 2DEG is located 150~nm underneath the crystal surface. Split gates, arranged as shown in the insets of Fig.~\ref{fig:3}, form a 90$^\circ$ corner with a QPC along each leg. One split gate is shared by both QPCs and defines the corner. Devices with a chamfered corner (Fig.~\ref{fig:3}a and b) as well as a straight corner (Fig.~\ref{fig:3} c) are investigated. The electron density $n$ equals 2.5$\cdot10^{11}$cm$^{-2}$ and 2.2$\cdot10^{11}$cm$^{-2}$ in the devices used in Fig.~\ref{fig:3}a,b, and c respectively. The wall length $a$ from each QPC to the corner is 3~$\mu$m. The electron mean free path of the 2DEG is 45~$\mu$m, one order of magnitude larger than the ballistic electron trajectories relevant for these studies. Transport measurements in a perpendicular magnetic field are carried out at 1.4~K by driving a sinusoidal current $I$ of 5~nA with a frequency of 13.3~Hz through the injector QPC. The gate voltage applied to all three gates defining the QPCs is identical. The voltage drop across the detector QPC, $V$, and the injector QPC are detected with separate voltage probes using a lock-in technique (measurement configuration shown in the insets of Fig.~\ref{fig:3}).

Three typical experimental data sets are plotted in Fig.~\ref{fig:3}.  The resistance data $R_\text{focusing}=V/I$ are recorded on devices which have an identical size but possess different realizations of the disorder. Since all devices are fabricated from the same heterostructure, the disorder is characterized by approximately the same statistical parameters. The curves within each panel are acquired for different gate voltage, i.e. different values of the QPC resistances. For the bottom curves at low QPC resistances many modes propagate \cite{L25,L26}. The top curves are recorded for transmission of a single mode. The collimation features at odd values of B/B$_0$ are in general much broader than the focusing features at even values in agreement with the broad distribution for the particle flow density in the left panel of Fig.~\ref{fig:2}c. They frequently split into two peaks depending on the device, i.e. the specific disorder realization (for instance at $B/B_0$=1 in panel b and at $B/B_0$=3 in panel c). We assert that the splitting of the collimation features is a result of disorder induced branching as in the left panel of Fig.~\ref{fig:2}c. Such splitting is absent for the focusing features at even values of $B/B_0$. The focusing features are more robust against variations of the disorder potential. The data in panel c were recorded on a device with a corner that is not chamfered. Processing as well as depletion will smoothen this corner somewhat however the direction of the specular reflection is not as well defined and focusing features which require specular reflection in this corner have dropped in amplitude.
\begin{figure}[t]
\includegraphics{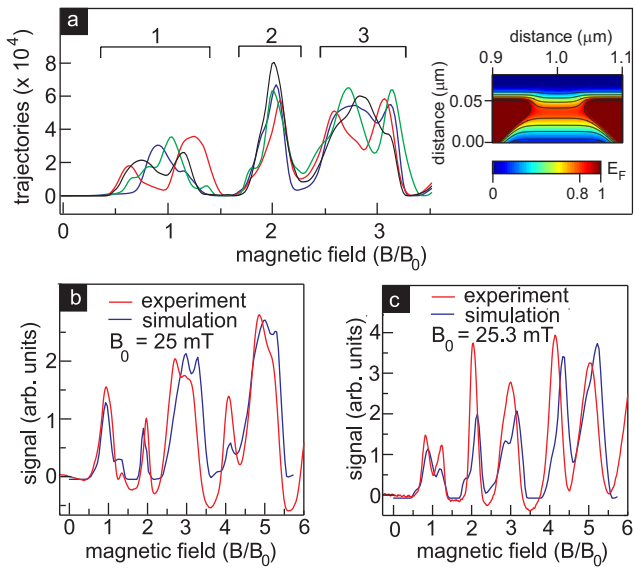}\caption{\label{fig:4}
Numerical transport simulation and its comparison with experimental results.(a) The number of trajectories which reach the collecting QPC as a function of $B/B_0$ in different realizations of the disorder potential in a corner device with chamfered corner. Peaks at odd multiples of $B_0$ are strongly affected by branching. The inset shows a contour plot of the potential used to simulate the QPC. (b-c) Comparison of experimental data (magnetotransport curves for $R_\text{QPC}$~$\approx$~3~k$\Omega$ in Figs.~\ref{fig:3}b and \ref{fig:3}c.) and numerical simulations for different disorder realizations.}
\end{figure}
\section{Transport simulation in the corner device}
To corroborate our assertion that the splitting of the collimation features comes from disorder induced branching, we have numerically calculated resistances for different disorder potential realizations. We simulate transport in the corner device by following classical trajectories from the emitting QPC until they either reach the collecting QPC and contribute to the transmission, or until they leave the system to the right of the collector. Within the Landauer-B\"{u}ttiker-formalism, we approximate the resistance $R_\text{focusing}$ measured in a four-terminal setup to be proportional to the transmission probability from emitter to collector. The QPC is tailored as a sum of variable-depth hyperbolic tangents, and we assume the presence of a saddle potential inside the QPC. Soft wall effects due to depletion are modeled by using a quadratic potential \cite{L30}. In the simulations, it is assumed that the particles enter the QPC from a lead with a cosinusoidal angular distribution. The saddle potential then has the effect of collimating the flow. We find that to obtain a collimation peak which is consistent with the experimental results, a saddle potential of approximately 80$\%$$E_F$ is needed. The precise functional form of the saddle potential is given in [\onlinecite{L30}], and is illustrated in the inset of Fig.~\ref{fig:4}. We note that the results presented here do not depend significantly on the parameters which describe the electrostatic walls. The weak disorder potential used in the simulations is modeled as a Gaussian random field with zero mean and standard deviation of the amplitude of the disorder potential V$_0$ = 2$\%$E$_F$, and a Gaussian correlation function $\langle V(\mathbf{r})V(\mathbf{r'})\rangle$~=~$V_0^2 e^{-\vert \mathbf{r}-\mathbf{r'}\vert^2/l_{cor}^2}$ , with correlation length $l_{cor}$ = 180~nm. As these parameters are not readily accessible in the experiment, they were chosen from a realistic range of values \cite{LL1}. The mean free path in this model potential is even larger than the measured mean free path, which is limited by other scattering processes, e.g. scattering by charged crystal defects. These are, however only relevant on length scales larger than our system size. Although our model system is even "deeper" in the ballistic regime than suggested by the experimentally evaluated mean free path, the weak disorder nevertheless has a pronounced effect on the peak structure. Figure~\ref{fig:4}a shows calculated resistance traces for different realizations of disorder potential with the same statistical parameters. We point out that the results are not sensitive to small variations in the parameters chosen for the random potential. Figures~\ref{fig:4}b and \ref{fig:4}c compare the experimental traces from Figs.~\ref{fig:3}b and \ref{fig:3}c with the calculated resistances. Finding potential disorder landscapes which produce such excellent agreement with experiment is to some extent accidental. 
\begin{figure}
\includegraphics{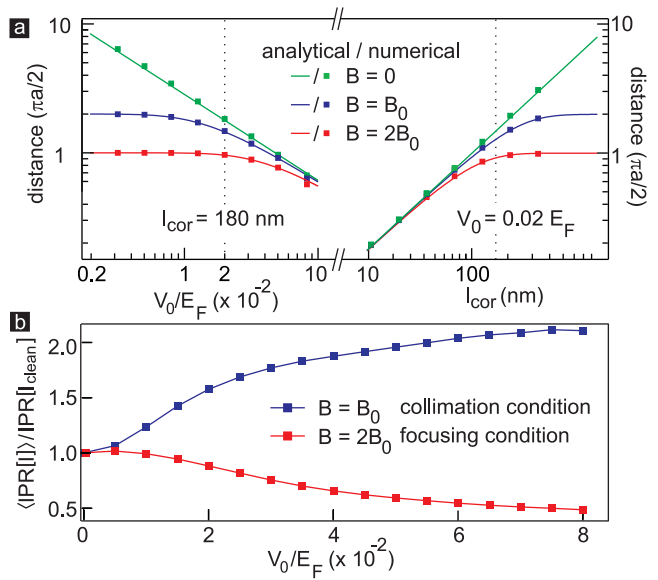}\caption{\label{fig:5}
Quantification of the influence of disorder. (a) The average distance an electron travels until a caustic forms as a function of the disorder parameters for $B~=~0$, $B~=~B_0$ and $B~=~2B_0$. The dotted lines mark the parameters of the random potential chosen for the simulations in Fig.~\ref{fig:4}. (b) The numerically calculated inverse participation ratio (IPR) of the flow density along the bottom boundary between $x$ = 0.5$a$ and $x$ = 1.5$a$, where $a$ is the distance from QPC to the corner, as a function of the standard deviation of the amplitude of the disorder potential under collimation and focusing conditions ($B=B_0$ and $B=2B_0$), averaged over 200 realizations of the random potential. The correlation length is identical to the one for the calculations in Fig.~\ref{fig:2}, and the IPR is normalized by the IPR of the clean system. The rising IPR at $B=B_0$ indicates that increasing disorder produces more peaks in the flow density, while the focusing peak at $B=2B_0$ is broadened by the disorder.}
\end{figure}
The main purpose of the simulations is to demonstrate that different disorder landscapes characterized by the same statistical parameters can indeed either lead to a pronounced splitting of collimation related peaks as one would expect from the behavior of the flow density at the sample boundary in the bottom left panel of Fig.~\ref{fig:2}d, or not. The focusing features at even values of $B/B_0$ do not show any splittings for the chosen parameters of the disorder. With decreasing correlation length or increasing $V_0$ however, we anticipate that also magnetic focusing features are affected by disorder induced branching. To assess the influence of the disorder as a function of the correlation length and $V_0$, it is instructive to calculate the average distance an electron travels until a caustic forms, $l_{caustic}$. We have obtained an analytical expression for this quantity, which is given by
\[
l_{caustic}=\pi^{2}\kappa\left(\textrm{Ai}\left[-\kappa^{2}B^{2}\right]^{2}+\textrm{Bi}\left[-\kappa^{2}B^{2}\right]^{2}\right)
\]
where Ai and Bi are Airy functions of the first and second kind and where $\kappa$ is a function of the disorder potential (see Appendix A). Our result is plotted in Fig.~\ref{fig:5}a for $B = 0$, $B_0$ and $2B_0$. For the latter field, magnetic focusing causes a caustic in our corner geometry after electrons have traveled on average a distance $\pi a/2$.  Indeed, $l_{caustic}$ saturates to this value in the limit of large correlation lengths and small $V_0$. For both values of the magnetic field we observe that for the parameters chosen in the simulations (indicated by dashed vertical lines) the mean distance to the first caustic starts to deviate from the case of magnetic focusing without disorder. For zero magnetic field $l_{caustic}$ scales like $l_{cor} /V_0^{2/3}$. \cite{L13,Kaplan2002,L27} Hence, branching can occur on much shorter length scales than the mean free path, which scales as $l_{cor}/V_0^2$. \cite{L28} Our analytical calculation shows that branching influences the transmission properties of our device. From Fig.~\ref{fig:5}a it is however not apparent that disorder has a different impact on collimation ($B = B_0$) than on focusing features ($B = 2 B_0$). To assess the impact of branching more quantitatively, we study the peakedness of the flow density along the bottom boundary of the corner device. Examples of the flow density $I(x)$ are displayed in Fig.~\ref{fig:2}c and d. As evident from this figure, the disorder potential induces multiple peaks near the bottom QPC at $B = B_0$, while the peakedness of the flow density for $B = 2B_0$ resembles that of the flow density in the absence of disorder. To capture the peakedness of the curves in a single quantity we use the inverse participation ratio IPR[I]=$\int^{x_2}_{x_1}dx I^2(x)/ (\int^{x_2}_{x_1}dx I(x))^2$. This inverse participation ratio in the vicinity of the bottom QPC is plotted as a function of the standard deviation of the disorder amplitude in Fig.~\ref{fig:5}b. For the collimation condition $B = B_0$, IPR[I] rapidly rises to higher values. It reflects the appearance of additional peaks. Under focusing conditions, IPR[I] drops indicating that the original peak mainly broadens. This confirms our experimental observations.

\section{Conclusion and Acknowledgment}
In conclusion, the weak disorder potential, which is present in any real two dimensional electron gas, causes a pronounced modification of collimation features due to branching. Our findings indicate that branching needs to be taken into account when interpreting transport data of mesoscopic devices even for state-of-the art heterostructures, for which the mean free path is an order of magnitude larger than the device size. 

DM and JJM contributed equally to this work. We acknowledge  technical support of Ulrike Waizmann and Monika Riek. This work was supported by the DFG research group 760 and the BMBF.

\appendix
\section{Mean distance to the first caustic in a random potential and magnetic field}
In order to determine the location of caustics, we consider an equation for the curvature of the action function $S$, which is obtained from the Hamilton-Jacobi-equation (HJE). Points along a trajectory where the curvature diverges indicate the position of a caustic. In a constant magnetic field $B$, the electron trajectories are circular. Considering small deviations from the circular paths in polar coordinates $\{r,\hat{\phi}\}$ allows a quasi-2D treatment, similar to the one of Fig.~2, in which time is identified with the angular variable as follows. The HJE in the symmetric gauge with vector potential $A~=~\frac{1}{2}r\hat{\phi}$ is given by
\[
\partial_tS +\frac{1}{2}(\partial_rS)^2+\frac{1}{2}r^2B^2+V(r(t))=0,
\]
where we have identified $t$ with $j$. Taking two derivatives with respect to $r$ , and evaluating the equation for the curvature $u=\partial_{rr}S(r)$ along the trajectories, we obtain the following equation for $u$:
\[
\frac{d}{dt}u+u^2+B^2+\partial_{rr}V(r)=0.
\]

For weak random potentials, we can approximate the correlation function of the random
potential as $c(r-r^{\prime},\phi-\phi^{\prime})=\left\langle V(r,\phi)V(r^{\prime}\phi^{\prime})\right\rangle=\delta(\phi-\phi^{\prime})A(r-r^{\prime})$. Thus, extending results from Refs.[\onlinecite{L27},\onlinecite{Klyatskin1993},\onlinecite{White1998}], we derive a Fokker-Planck-equation for the probability density  $p(t,u)$
\[
\frac{d}{dt}p(t,u)=\frac{\partial}{\partial u}(u^2+B^2)p(t,u)+\frac{D}{2}\frac{{\partial}^2}{\partial u^2}p(u,t),
\]
where $D=\int^{\infty}_{-\infty}\frac{{\partial}^4}{\partial y^4}c(x,y)|_{y=0} dx$. For a Gaussian correlation function
\[
c(x,y)=V_0^2 e^{-(x^2+y^2)/l^2_{cor}} 
\]
we obtain $D=12\sqrt{\pi}V_0^2l_{cor}^{-3}$.
To obtain an equation for the onset of the branching, we now derive an expression for the mean time it takes for a caustic to develop along a trajectory. This can be done treating the problem as a mean first passage time problem [\onlinecite{L30},\onlinecite{Risken1989}]. From this expression, one can then easily derive the mean distance traveled along a trajectory until a caustic is hit. Of greatest importance for the experiment is the point source with initial condition $u_0=\infty$. The corresponding mean distance to the first caustic $l_{caustic}$ is then calculated [\onlinecite{L30}] to be
\[
l_{caustic}=\pi^{2}\kappa\left(\textrm{Ai}\left[-\kappa^{2}B^{2}\right]^{2}+\textrm{Bi}\left[-\kappa^{2}B^{2}\right]^{2}\right)
\]
where Ai and Bi are Airy functions of the first and second kind [\onlinecite{Abramowitz1972}], and where $\kappa = (2/D)^{1/3}$.

%

\end{document}